%% file: example_paper.tex
\documentclass{article}

\usepackage{microtype}
\usepackage{graphicx}
\usepackage{subcaption}
\usepackage{booktabs} 
\usepackage{balance}
\usepackage{tabularx}
\usepackage{hyperref}

\usepackage[preprint]{icml2026}

\usepackage{amsmath}
\usepackage{amssymb}
\usepackage{mathtools}
\usepackage{amsthm}
\usepackage{soul, xcolor}

\usepackage[capitalize,noabbrev]{cleveref}

\theoremstyle{plain}

\theoremstyle{definition}

\theoremstyle{remark}

\usepackage[textsize=tiny]{todonotes}

\icmltitlerunning{Rethinking Continual Learning for Speech and Audio: A Representation-Centric Taxonomy and Open Problems}

\begin{document}

\twocolumn[
    \icmltitle{Rethinking Continual Learning for Speech and Audio: A Representation-Centric Taxonomy 
    and Open Problems}



  \icmlsetsymbol{equal}{*}

  \begin{icmlauthorlist}
    \icmlauthor{Yang Xiao}{yyy}
    \icmlauthor{Siyi Wang}{yyy}
    \icmlauthor{Eun-Jung Holden}{yyy}
    \icmlauthor{Ting Dang}{yyy}
  \end{icmlauthorlist}

  \icmlaffiliation{yyy}{University of Melbourne, Melbourne, Australia}

  \icmlcorrespondingauthor{Ting Dang}{ting.dang@unimelb.edu.au}

  \icmlkeywords{Machine Learning, ICML}

  \vskip 0.3in
]



\printAffiliationsAndNotice{}  

\begin{abstract}
Speech and audio systems operate in inherently non-stationary environments, yet continual learning (CL) research in this domain, especially in the foundation model era, remains fragmented that fail to account for the coupled, geometry-sensitive nature of acoustic representations. Modern large audio language models (LALM) operate over highly entangled, continuous representations that jointly encode linguistic, speaker, and paralinguistic factors within a shared latent space. CL is therefore fundamentally about preserving and evolving shared representation structure rather than retaining isolated task knowledge. In this work, we revisit CL for speech from a representation-centered perspective, and introduce a new taxonomy that organizes CL according to how underlying representation geometry evolves under non-stationary acoustic conditions. We further identify key mismatches between current CL assumptions and LALMs behavior, and finally outline a set of open challenges and future research directions. 

\end{abstract}

\section{Introduction}

The real world is inherently continuous and non-stationary. Acoustic environments evolve, speakers age, languages and accents shift, and novel sound events emerge over time. However, most speech systems are still trained once on static datasets and deployed under the assumption that the underlying data distribution remains stable. This mismatch between the dynamic nature of acoustic experience and the static paradigm of model training is precisely the setting that continual learning (CL) aims to address~\cite{de2021continual}. Yet existing CL methodologies were largely developed under assumptions that align poorly with speech and audio data: discrete task boundaries, stationary input distributions, and relatively disentangled representations learned from small-scale models trained from scratch. As a result, many established CL techniques encounter fundamental limitations when applied to modern speech systems.

Speech presents challenges that differ fundamentally from vision and text due to both the nature of the acoustic signal and the structure of its learned representations. \emph{At the signal level}, vision and text operate on relatively stable and explicit units: pixels are spatially localized and tokens are discretized symbolic units with predefined boundaries. In contrast, speech is a continuous temporal waveform whose meaningful structures, such as phonemes, words, speaker turns, and prosodic events, are not explicitly given but must be inferred by the model itself. \emph{At the representation level}, speech embeddings simultaneously encode multiple highly entangled factors, including linguistic content, speaker identity, emotion, accent, recording conditions, and paralinguistic information. Unlike many vision or NLP settings where task-relevant attributes can often be partially separated or localized, these factors in speech are deeply coupled within a shared latent geometry. Consequently, adapting the model for one objective frequently perturbs representations required for others~\cite{chen2025aft}. For example, continual adaptation for emotion recognition may unintentionally distort phonetic structures important for automatic speech recognition (ASR) or speaker characteristics required for speaker verification. Continual learning in speech therefore cannot be framed simply as preserving task-specific knowledge; rather, it requires maintaining the stability of a shared and continuously evolving representational geometry across competing objectives and acoustic conditions.

These challenges become even more pronounced in the era of speech and audio foundation models. Recent years have witnessed a rapid shift from task-specific architectures toward large-scale pretrained models such as \textit{wav2vec 2.0}~\cite{Baevski2020wav2vec2}, \textit{HuBERT}~\cite{Hsu2021HuBERT}, \textit{Whisper}~\cite{Radford2022Whisper}, and recently LALMs~\cite{yang2025towards}. Trained on massive and heterogeneous audio corpora, these models learn highly generalizable latent representations that support a broad spectrum of capabilities, including ASR, speaker verification, emotion recognition, audio captioning, spoken dialogue, and multimodal reasoning. Consequently, continual learning is no longer simply a problem of sequentially acquiring new tasks. Instead, it becomes a problem of continuously adapting a shared pretrained representation while preserving the acoustic and linguistic structures that underpin diverse downstream capabilities. Under this setting, catastrophic forgetting can manifest not only as performance degradation on previously learned tasks, but also as gradual corruption of the latent representational geometry itself. A model may maintain benchmark-level accuracy while its internal structure progressively deteriorates, for example through reduced phonetic separability, compressed speaker manifolds, or degradation of paralinguistic organization. Such representational shifts are particularly problematic in speech because multiple capabilities depend on the same shared embedding space.

Existing adaptation strategies only partially address this challenge. Parameter-efficient tuning approaches, including LoRA~\cite{hu2022lora}, adapters~\cite{Sel23}, and prompt-based adaptation~\cite{Cui2025AAAI}, reduce the extent of destructive full-model updates, but they do not eliminate representational drift and may introduce new forms of interference across tasks, domains, and adaptation stages. As foundation models continue to expand in scale and capability, the central question of continual learning for speech shifts from \emph{how to prevent forgetting previously learned tasks} to \emph{how to continuously adapt foundation-scale speech representations while preserving the latent structures that support broad acoustic and linguistic generalization}.


We argue that the combination of speech-specific representational complexity and foundation-scale pretrained models fundamentally reshapes the continual learning problem beyond the conventional CL settings. This paper presents a representation-centered perspective on CL for speech and audio in the foundation model era. We introduce a taxonomy of speech-centric continual learning scenarios, analyze how existing CL paradigms fail under the dynamics of large-scale speech representations and adaptation, and outline future research directions centered on preserving representational geometry, adapting entangled latent spaces, and developing evaluation protocols that reflect the behavior and degradation modes of modern large audio language models (LALMs).

\input{02-taxnomy}

\section{The Evolution of Mitigation Strategies}
This section examines how the existing methods\footnotemark have evolved and why the coupled nature of modern speech representations presents a primary bottleneck for continuous adaptation. We categorize existing approaches into three fundamental mechanisms: replay, regularization, and architectural isolation. 

\footnotetext{Full references are in our \href{https://github.com/swagshaw/Awesome-Speech-and-Audio-Continual-Learning}{GitHub list}.}

\noindent\textbf{Replay-Based Methods. }
Replay strategies mitigate forgetting by revisiting previous data distributions. In early stages of model development, replaying raw waveforms or acoustic features was standard practice. Interleaving old audio samples with new data anchored the model during the acquisition of new tasks, such as automatic speech recognition~\cite{Cha21}, multilingual speech synthesis~\cite{Yan21}, and audio classification~\cite{xiao2022continual,Xia22,Pen24,Xia24,Xia25b}. This approach is effective because directly replaying raw waveforms preserves the full geometric space. However, as models scale, speech data encounters strict privacy constraints and storage limits. 

\noindent\textbf{Regularization-Based Methods. }
Regularization methods impose soft constraints on model updates to preserve previously learned knowledge without requiring access to old data. Traditional methods, such as Elastic Weight Consolidation (EWC)~\cite{Kirkpatrick2017EWC} and Learning without Forgetting (LwF)~\cite{Li2016LwF}, were widely applied to early ASR models~\cite{Gho19,Hou20}. By computing Fisher information matrices, these approaches identified and penalized changes to weights essential for previous tasks. These approaches can be viewed as indirect mechanisms for geometry preservation, enforcing stability at the parameter level rather than directly constraining the latent representation space. However, this assumption becomes increasingly fragile in modern LALMs, where representations are highly entangled across phonetic, speaker, and acoustic factors. In such settings, parameter-level constraints are insufficient to prevent representational drift: acoustic conditions and phonetic structure are jointly encoded, so even small updates can induce global distortions in embedding geometry. 


\noindent\textbf{Architectural Isolation Methods.} 
Architectural isolation tackles shared module interference by freezing the main pretrained backbone and updating only task-specific, lightweight parameters. Previously research utilize adapter to constrain the trainable parameters~\cite{Xia24b}. With the introduction of foundation models like wav2vec 2.0~\cite{Baevski2020wav2vec2} and Whisper~\cite{Radford2022Whisper}, PEFT has become widely adopted. For example, integrating LoRA~\cite{hu2022lora} modules into Whisper allows the model to learn new languages sequentially without losing original capabilities~\cite{Xu24,Liu24,Yue25,Xia26,xiao2026adapting}. PEFT-based methods can be interpreted as mechanisms for geometry preservation through parameter isolation. However, this assumption breaks down in speech because representational entanglement is not aligned with parameter modularity. Unlike NLP backbones, audio encoders encode strong low-level spectral structure, where speaker identity, phonetics, and paralinguistic cues are jointly embedded within the same continuous acoustic manifold~\cite{li2026pacepretrainedaudiocontinual}. As a result, isolating updates to bottleneck modules does not isolate their effect on the representation geometry. 


\vspace{-5pt}
\section{LALMs Post-Training as Implicit CL}

\begin{figure}
    \centering
    
\includegraphics[width=\linewidth]{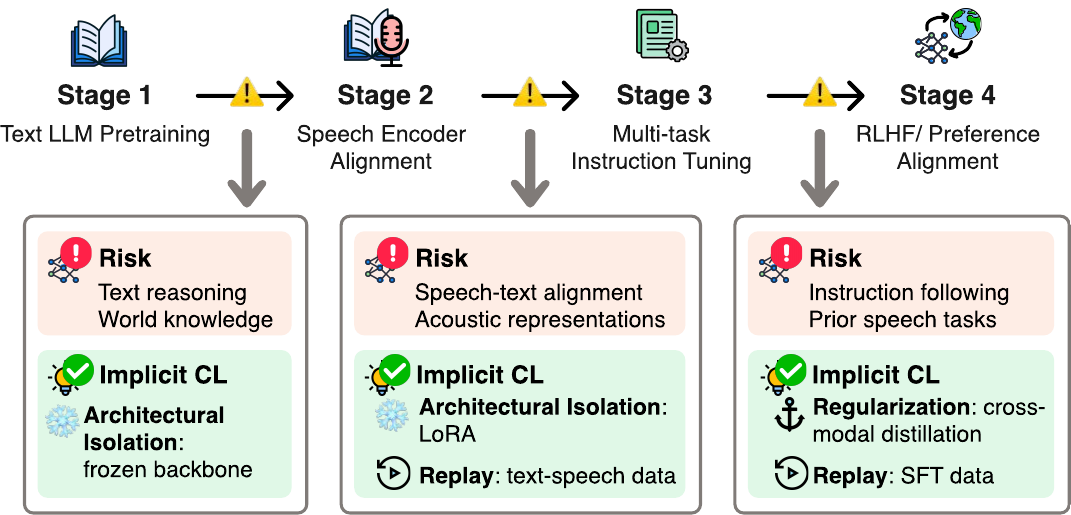}
\vspace{-8pt}
    \caption{Decoding Speech LLM Post-Training as an Implicit Multimodal Continual Learning Pipeline. The 4-stage development process (from text-only pretraining to preference alignment). }
    \label{fig:placeholder}
    \vspace{-24pt}
\end{figure}

When the representation-centric perspective introduced above is applied to the LLM era, it indicates an underexplored reality: much of the current LALM engineering practice already implements implicit continual learning. The standard multi-stage post-training pipeline is, in fact, a sequence of complex, cross-modality continual learning problems. So we can map standard LALM recipes directly to their underlying CL mechanisms as illustrated in Figure 1:

\textbf{Stage 1 → 2 (Speech Encoder Alignment):} When transitioning from a pretrained text LLM to cross-modal alignment, the model is at risk of forgetting its existed text reasoning and world knowledge. To avoid this, researchers almost universally rely on freezing the text backbone while training a speech encoder, which is fundamentally architectural isolation~\cite{Hsi25,Cue25}.

\textbf{Stage 2 → 3 (Multi-task Instruction Tuning): }As the model learns to follow diverse audio instructions, the speech-text alignment and acoustic representations established in Stage 2 are at risk. To anchor the embedding space, the community typically mixes text and speech instruction data which acting as replay\cite{chu2024qwen2} while using LoRA and adapters to constrain the update space as architectural isolation\cite{Xu24,Liu24}.

\textbf{Stage 3 → 4 (RLHF / Preference Alignment): }In the final alignment phase, the model is at risk of losing its prior instruction like following behavior and legacy speech tasks (e.g., ASR, TTS). To prevent this, engineers often inject cross-modal distillation to explicitly transfer prior competence~\cite{Wan25} alongside data replay. Recent theoretical insights propose that the very use of on-policy Reinforcement Learning (RL) in this stage also acts as a fundamental implicit CL strategy~\cite{shenfeldrl}. On-policy RL biases model updates toward KL-minimal solutions relative to the base policy. By minimizing this distributional shift, RL naturally mitigates catastrophic forgetting and preserves prior multimodal capabilities significantly better than standard offline supervised fine-tuning.

When reviewing these transitions, a clear trend appears: LALM post-training is the only area where hybrid methods are the norm rather than the exception. Since the model must maintain multiple abilities at once (such as text reasoning, speech-text alignment, prior speech tasks, and instruction-following behavior), no single continual learning approach is enough. The field is increasingly relying on combined methods in practice. For instance, maintaining performance often requires keeping text backbones frozen, replaying text data, and using cross-modal distillation at the same time~\cite{Hsi25,Cue25,wang2025continualspeechlearningfused,Wan26}. This practical consensus has formed even without a complete theoretical explanation.


\vspace{-5pt}
\section{Open Problems and Future Directions}
The transition from mitigating task-specific forgetting to preserving continuous representational geometry presents a distinct set of unresolved challenges. 

To advance beyond standard continual fine-tuning, developing scalable continual pre-training mechanisms is necessary~\cite{roth2024practitioner}. As discussed in our analysis of replay-based methods, a primary constraint in the speech domain is the privacy of biometric voice data, which limits the use of raw-audio memory buffers. Because speaker identity and linguistic content are tightly coupled at the perception layer, storing raw data risks biometric exposure. Transitioning from explicit external rehearsal to internal generative pseudo-replay offers a structural solution~\cite{frascaroli2024clip}. LALMs can be designed to self-generate modality-aligned pseudo-replays directly from their internal latent space. This strategy bypasses the need for raw data storage while maintaining the geometric equilibrium of the acoustic manifold during continuous updates.

Real-world speech systems frequently operate under dynamic conditions, including missing text metadata or corrupted audio streams. Existing multimodal continual learning methods typically assume the consistent availability of all modalities~\cite{huang2025mind}. When applied to speech, this assumption is problematic: relying on a dominant modality during incomplete updates distorts the cross-modal alignment at the semantic layer. Addressing this requires architectures capable of generalizing under missing modalities without disrupting established phonetic boundaries. Future research should model this modality gap directly within the shared embedding space. Implementing dynamic routing or masking mechanisms can mathematically protect cross-modal alignment from the feature drift.

\clearpage
\balance
\bibliography{example_paper}
\bibliographystyle{icml2026}




\end{document}

%% file: 02-taxnomy.tex
\section{A Representation-Centric Taxonomy of Continual Learning in Speech}
Classical CL taxonomies~\cite{de2021continual} typically organize learning scenarios according to shifts in tasks, domains, or label spaces, leading to well-established settings such as task-incremental, domain-incremental, and class-incremental learning. While effective in controlled benchmarks, these formulations provide only a partial view of continual learning in modern speech systems. 
We therefore propose a representation-centric taxonomy of continual learning in speech and audio, characterizing continual learning according to how the underlying representation evolves over time and which structural properties must be preserved during adaptation. We identify four forms of representational evolution: geometry preservation, expansion, alignment, and specialization.

\noindent\textbf{Geometry Preservation.}
Geometry preservation refers to settings where the primary objective is to maintain existing representational structure under distributional shift. In real-world deployment, speech systems encounter continuously changing acoustic conditions such as new speakers, recording devices, noise profiles, and channel effects. While adaptation improves robustness, repeated updates can gradually distort previously learned latent geometry, leading to effects such as reduced phonetic separability, collapsed speaker manifolds, or weakened paralinguistic structure. The goal of geometry preservation is therefore to constrain adaptation such that existing structure remains stable while allowing limited adjustment to new input distributions.

\noindent\textbf{Geometry Expansion.}
Geometry expansion describes scenarios in which the representation must incorporate previously unseen information while preserving compatibility with existing structure. This includes new languages, accents, vocabularies, speakers, and acoustic events. The central challenge is balancing plasticity and stability within a shared latent space: new information must be embedded without overwriting or fragmenting previously established organization. For example, multilingual extension requires integrating new phonetic systems without degrading separability of existing languages, while speaker expansion requires adding new identity structure without collapsing previously learned distinctions.

\noindent\textbf{Geometry Alignment.}
Geometry alignment refers to CL settings in which relationships between multiple representation spaces must be preserved or updated consistently. This is particularly important in speech foundation models that integrate acoustic encoders with language models, multimodal modules, or external memory systems. Continual adaptation may introduce alignment drift, where mappings between speech representations and textual or multimodal spaces degrade even when individual modalities remain stable. For instance, updating a speech encoder may break its correspondence with a frozen language model, leading to degraded speech-to-text performance.

\noindent\textbf{Geometry Specialization.}
Geometry specialization captures settings where the representation is adapted to support new or refined capabilities built on a shared foundation model. These include audio captioning, spoken question answering, and agentic dialogue. Adaptation typically reshapes or reweights existing regions of the representation space to emphasize task-relevant structure, improving performance on new objectives but potentially interfering with previously learned capabilities that rely on overlapping representations. This highlights the tension between capability acquisition and representational reuse in foundation models.

\vspace{-5pt}
\paragraph{Adaptation Perspective.}
While the taxonomy above characterizes what changes in the representation space, CL behavior is also determined by where these changes are induced within the model. In modern speech foundation models, adaptation occurs across multiple components, including \emph{acoustic encoders, alignment modules, language models, memory systems, and agentic components}. These layers provide a complementary mechanistic view, as the same form of representational evolution may arise from different adaptation sites, leading to distinct interference patterns and forgetting behaviors.

Acoustic encoder adaptation primarily affects low-level acoustic geometry and is most closely associated with geometry preservation and expansion, as it directly shapes phonetic structure, speaker identity, and environmental robustness. Alignment-layer adaptation governs cross-modal correspondence and is central to geometry alignment, ensuring stable mappings between speech and textual or multimodal representations. Language-model adaptation influences higher-level semantic reasoning and is often linked to geometry specialization. Memory systems support incremental knowledge accumulation and user-specific information, making them closely related to geometry expansion. In contrast, agent-level adaptation operates at the level of behavioral policy and interaction, corresponding primarily to geometry specialization through changes in planning, tool use, and decision-making. 

\vspace{-5pt}
\paragraph{Discussion.}
This taxonomy reframes CL in speech from \emph{the interaction between evolving representations and the architectural layers through which adaptation is performed}. Representation evolution and adaptation location are only partially coupled: the same form of geometric change can be induced by updates at different layers, and updates to a single layer may simultaneously impact multiple forms of representational geometry. 
These forms of evolution are not mutually exclusive and often co-occur in real systems, providing a unified perspective that connects representation dynamics with system-level adaptation mechanisms in modern speech foundation models.